\documentstyle[11pt,paspconf]{article}

\begin{document}

\title{Stellar Mass Function From SIM Astrometry/Photometry}
\author{A.\ Gould and S.\ Salim}
\affil{Dept.\ of Astronomy, Ohio State University, 174 W. 18th Ave.,
Columbus, OH 43210}

\begin{abstract}

\end{abstract}

By combining SIM observations with ground-based photometry, one can
completely solve microlensing events seen toward the Galactic bulge.  One
could measure the mass, distance, and transverse velocity of $\sim 100$ lenses 
to $\sim 5\%$ precision in only $\sim 500$ hours of SIM time.  Among the
numerous applications are 1) measurement of the mass functions (MFs) of the
bulge and disk 2) measurement of the relative normalizations of the bulge
and disk MFs (and so their relative contribution to the Galactic potential),
3) measurement of the number of bulge white dwarfs and neutron stars
(and so the initial MF well above the present turnoff).  SIM astrometric
measurements are simultaneously photometric measurements.  SIM astrometry
determines the angular size of the Einstein ring on the sky, and comparison
of SIM and ground-based photometry determines the size of the Einstein
ring projected onto the observer plane.  Only by combining both of these
measurements is it possible to completely solve the microlensing events.


\section{Introduction}

	Microlensing observations toward the Galactic bulge are yielding
important clues about the structure of the Milky Way (Udalski et al.\ 1994;
Alcock et al.\ 1997).  However, the only useful parameter that is usually
extracted from a microlensing event is its timescale, $t_e$, which is
a complicated combination of the three parameters one would like to 
know about the lens, its mass $M$, its distance $d_l$, and its proper motion
relative to the observer-source line of sight $\mu$.  Specifically,
\begin{equation}
t_e = {\theta_e\over \mu},\qquad \theta_e = \sqrt{4 G M\over c^2 D}, 
\label{eqn:tedef}
\end{equation}
where $\theta_e$ is the angular Einstein radius (the characteristic
angular size over which the lens has a significant effect),
\begin{equation}
D\equiv {d_l d_s\over d_s-d_l},
\label{eqn:Ddef}
\end{equation}
and $d_s$ is the distance to the source.  Thus, if one wants to use 
microlensing observations, for example, to measure the bulge mass function
(MF),
one can analyze the distribution of timescales (Zhao, Spergel, \& Rich 1995;
Han \& Gould 1996), but to do so one must make a whole series of 
model-dependent assumptions, such as the distributions
of the source-lens relative velocities, the source distances, and the
lens distances, and
the proportion of events that are due to foreground lenses in the disk
rather than in the bulge itself.

	The scientific return from bulge microlensing observations would
be increased many fold if it were possible to measure $M$, $D$, and
$\mu$, separately for each event, especially if this were combined with
measurements of $d_s$ and $\mu_s$, the distance and proper motion of the
source.  With these additional pieces of information, one could determine
both $d_l$ and the absolute transverse velocity of the lens.  

	First, bulge and disk lenses could be separately 
identified (from their distances and kinematics) so that the bulge
and disk MFs could be measured separately and unambiguously.
Second, the relative normalizations of the bulge and disk MFs could be
determined so that one would know how much of the Galactic potential
was attributable to each structure.  

	Third, it would be possible to
measure the number of white dwarfs and neutron stars in the bulge  
(Gould 1999). These
stars are substantially too faint to be detected optically in the
crowded bulge fields, but they would be easily revealed in a census of
{\it masses} of bulge microlensing events.  White dwarfs would show up
as a spike in the MF at $M\sim 0.6\,M_\odot$, and neutron stars have masses
that are higher than those of turnoff stars.  Note that the sharp white
dwarf feature in the MF is spread out to a fractional width of ${\cal O}(1)$
in the $t_e$ distribution, so the white dwarfs cannot be picked out from
the timescales.  The same is basically true of neutron stars since they
are only $\sim 1.5$ times heavier than turnoff stars.  Since white dwarfs
and neutron stars are remnants of main-sequence stars with masses
respectively $M_\odot < M_{\rm ms} < 8\,M_\odot$ and $M_{\rm ms}> 8\,M_\odot$,
the specific frequency of these remnants would in turn yield information about
the initial MF to very high masses.

	Fourth, one could determine whether the bulge contains massive
objects other than those associated with the observed stars.  A very 
puzzling (but often overlooked) fact is that kinematic studies of ellipticals
and spiral bulges typically yield mass-to-light ratios $M/L_V\sim 10 h\sim 7$,
much higher than the only two populations for which we have unambiguous
measurements:  dymanical studies of globular clusters yield
$M/L_V\sim 2$--3 (Pryor \& Meylan 1993), and a complete census of stars
in the disk (Gould, Bahcall, \& Flynn 1997) combined with a surface brightness
of the disk (Binney \& Tremaine 1987) yield $M/L_V\sim 2$.  It is usually
assumed that the bulge MF differs dramatically from these other populations.
However, it is quite possible that the bulge contains substantial quantities
of dark matter, either in compact objects or in diffuse material (WIMPs).
The MF of the
luminous stars in the bulge has now been measured in both the optical
(Holtzman et al.\ 1998) and the infrared (Zoccali et al.\ 1999), so that
if the total MF were measured from microlensing it would be possible
to distinguish among these various competing scenarios.

\section{Decoding Microlensing Events with SIM}

	SIM observations can completely solve for the physical parameters
of the lens and source, $M$, $D$, $\mu$, $d_s$, and $\mu_s$, by combining
two seemingly unrelated ideas:  Boden, Shao, \& Van Buren (1998) showed
that it was possible to measure $\theta_e$ from astrometric measurements
of the apparent source position.  Gould (1995) showed that it was
possible to measured the projected Einstein radius $\tilde r_e$ 
\begin{equation}
\tilde r_e = D\theta_e
\label{eqn:retildedef}
\end{equation}
from photometric measurements of the event simultaneously from the Earth
and a satellite in solar orbit.  It is clear that if both $\theta_e$ and
$\tilde r_e$ are measured, then one can measure $M$, $D$, and $\mu$.
\begin{equation}
M={c^2\over 4 G}\tilde r_e\theta_e,\qquad D={\tilde r_e\over\theta_e},
\qquad \mu = {\theta_e\over t_e}.
\label{eqn:threeevals}
\end{equation}
Moreover, as we will show below, in the course of measuring $\theta_e$ 
astrometrically, one automatically measures $\mu_s$ and $d_s$.  Thus, if
SIM can really carry out these two measurements simultaneously, 
bulge microlensing events can be completely solved.  How does this work?

\subsection{Astrometry}

	Suppose that a lens and source are separated on the sky by
${\bf u}\theta_e$, where ${\bf u}=(\tau,\beta)$ is the separation in
units of the Einstein radius, $\beta$ is the impact parameter of the event,
$\tau = (t-t_0)/t_e$, and $t_0$ is the time of closest approach.  Then
the source will be split into two images with positions ${\bf u}_\pm\theta_e$ 
and magnifications $A_\pm$,
\begin{equation}
{\bf u}_\pm = \biggl[{u\pm \sqrt{u^2+4}\over 2}\biggr]{{\bf u}\over u},
\qquad A_\pm = {A\pm 1\over 2},\qquad
A={u^2+2\over u\sqrt{u^2+4}}.
\label{eqn:adefs}
\end{equation}
The separation between the images $(\sim 2\theta_e)$ is of order 100s of
$\mu$as and so is far too small to be resolved by SIM with its 10 mas central
fringe.  However, as Boden et al. (1998) showed, the displacement of the
image centroid from the ``true'' position of the source is
\begin{equation}
(A_+{\bf u}_+ + A_-{\bf u}_- - {\bf u})\theta_e = {{\bf u}\over u^2+2}
\theta_e,
\label{eqn:deltau}
\end{equation}
and therefore has a maximum of $\theta_e/\sqrt{8}$ (at $u=\sqrt{2}$) and so is
well within SIM's capabilities.  

	Of course, just measuring the apparent position of the source
does not by itself yield the displacement due to lensing.  One must also
know where the source would have appeared in the absence of lensing.
To detemine this, one must measure the distance $d_s$ (i.e.
the parallax $\pi_s= {\rm AU}/d_s$) and proper motion 
$\mu_s$ of the
source at late times (when its apparent position is not influenced by the
lens), then project its ``true'' position backwards to the time of the
event.

	In principle, it is also possible to measure $\tilde r_e$ from
astrometry, but since the deviations caused by Earth's motion are a
higher order effect, this is not the most practical method
(Gould \& Salim 1999).

\subsection{Photometry}

	The Einstein radius projected onto the plane of the observer is
typically a few AU, and so the satellite and the Earth see significantly
different events, with different impact parameters $\beta$ and different times
of maximum $t_0$.  (The timescales $t_e$ are also slightly different,
but this is a higher order effect which we will ignore for the moment.)\ \
Hence, the position in the Einstein ring will differ by
\begin{equation}
\Delta {\bf u} = (\Delta \tau,\Delta \beta)
\label{eqn:deltaudef}
\end{equation}
where $\Delta \tau=(-t_{0,\rm sat}+t_{0,\oplus})/t_e$ and 
$\Delta \beta=\beta_{\rm sat}-\beta_\oplus$.
By measuring $\beta$ (from the peak magnification) and $t_0$ (from the
time of peak magnification) from the Earth and satellite, one can therefore
measure $\Delta \bf u$.  It is then possible to determine  $\tilde r_e$ by
using
\begin{equation}
\tilde r_e = {d_{\rm sat}\over \Delta u},
\label{eqn:reeval}
\end{equation}
where $d_{\rm sat}$ is the distance to the satellite projected onto the
plane of the sky.

	Actually, there is a bit of a complication in that the impact
parameter can be on either side of the lens so that $\beta_{\rm sat}$ and
$\beta_\oplus$ can each be of either sign, while the measurement of $\beta$
from the light curve is sensitive only to its square (i.e., its amplitude
but not its sign) (Refsdal 1966,; Gould 1994).  Hence
$\Delta\beta = \pm (\beta_{\rm sat}\pm \beta_\oplus)$ and so cannot be
unambiguously determined simply by measuring $\beta_{\rm sat}$ and
$\beta_\oplus$.
However, Gould (1995) showed that
this ambiguity could be resolved using the small difference in $t_e$ as
measured by the Earth and satellite.

\section{SIM: Simultaneous Astrometry and Photometry}

	In fact, although SIM is designed to do astrometry, the astrometric
measurements are done by counting photons over the central fringe.  The
sum of these photon counts is a photometric measurement.  Thus SIM
simultaneously does astrometry and photometry.  Of course, for most
purposes, photometry using 25 cm mirrors is not very interesting.
However, in the present case, the fact that SIM is making the photometric
measurement at several tenths of an AU from the Earth is what is crucial.
 For photon-limited measurements, the ratio of the fractional photometric 
error $\sigma_{\rm ph}$ to the astrometric error $\sigma_\theta$
is given by
\begin{equation}
\sigma_{\rm ph} = {\sigma_\theta\over \theta_f},\qquad \theta_f \equiv
{\lambda\over 2\pi d}\sim 2.5\,\rm mas,
\label{eqn:sigmaph}
\end{equation}
where $d\sim 10\,$m is the distance between the mirrors, $\lambda$ is the
wavelength of the light, in this cases taken to be $\lambda\sim 0.8\,\mu$m,
appropriate for bulge clump giants.

	Gould \& Salim (1999) showed that by combining such photometric
measurements that can be generated simultaneously with SIM astrometric
measurements, it should be possible to measure $M$, $D$, $\mu$, $d_s$, and
$\mu_s$ all to better than 5\% precision with about 5 hours of SIM time
for bulge microlensing events with $I=15$ sources.  Over the SIM
lifetime, there should be of order 100 such events, so about 100 mass
measurements are possible.

\acknowledgments

This work was supported in part by grant AST 97-27520 from the NSF and in
part by grant NAG5-3111 from NASA.

\end{document}